\documentclass[preprint,
groupedaddress,
superscriptaddress,
 amsmath,amssymb,
 aps,
 prl,
 showpacs
]{revtex4-2}

\usepackage{graphicx}
\usepackage{dcolumn}
\usepackage{bm}
\usepackage{xcolor}
\usepackage{epstopdf}
\AppendGraphicsExtensions{.tif}
\usepackage{graphicx}
\usepackage{dcolumn}
\usepackage{bm}
\usepackage{hyperref}
\hypersetup{
    colorlinks=true,
    urlcolor= blue,
    citecolor=blue,
linkcolor= blue}

\newcommand{\beginsupplement}{%
        \setcounter{table}{0}
        \renewcommand{\thetable}{S\arabic{table}}%
        \setcounter{figure}{0}
        \renewcommand{\thefigure}{S\arabic{figure}}%
     }
     
\begin{document}

\title{Field-tunable interactions and frustration in underlayer-mediated artificial spin ice}

\author{Susan Kempinger}
\thanks{Corresponding author: sekempinger@noctrl.edu}
 \affiliation{Department of Physics, The Pennsylvania State University, University Park, Pennsylvania 16802-6300, USA}
 \affiliation{Department of Physics, North Central College, Naperville, Illinois 60540, USA}
 \author{Yu-Sheng Huang}
 \affiliation{Department of Physics, The Pennsylvania State University, University Park, Pennsylvania 16802-6300, USA}
 \author{Paul Lammert}
 \affiliation{Department of Physics, The Pennsylvania State University, University Park, Pennsylvania 16802-6300, USA}
\author{Michael Vogel}
 \affiliation{Institute of Physics and Center for Interdisciplinary Nanostrucre Science and Technology (CINSaT), University of Kalle, Heinrich-Platt-Str. 40, 34132 Kassel, Germany}
 \author{Axel Hoffmann}
 \affiliation{Materials Research Laboratory and Department of Materials Science and Engineering, University of Illinois at Urbana-Champaign, Urbana, Illinois 61801, USA}
 \author{Vincent H. Crespi}
 \affiliation{Department of Physics, The Pennsylvania State University, University Park, Pennsylvania 16802-6300, USA}
 \author{Peter Schiffer}
 \affiliation{Department of Applied Physics and Department of Physics, Yale University, New Haven, CT 06520  USA}
 \author{Nitin Samarth}
 \thanks{Corresponding author: nsamarth@psu.edu}
 \affiliation{Department of Physics, The Pennsylvania State University, University Park, Pennsylvania 16802-6300, USA}

\date{\today}

\begin{abstract}
  
Artificial spin ice systems have opened experimental windows into a range of model magnetic systems through the control of interactions among nanomagnet moments.  This control has previously been enabled by altering the nanomagnet size and the geometry of their placement.  Here we demonstrate that the interactions in  artificial spin ice can be further controlled by including a soft ferromagnetic underlayer below the moments. Such a substrate also breaks the symmetry in the array when magnetized, introducing a directional component to the correlations. Using spatially resolved magneto-optical Kerr effect microscopy to image the demagnetized ground states, we show that the correlation of the demagnetized states depends on the direction of underlayer magnetization. Further, the relative interaction strength of nearest and next-nearest neighbors varies significantly with the array geometry. We exploit this feature to induce frustration in an inherently unfrustrated square lattice geometry, demonstrating new possibilities for effective geometries in two dimensional nanomagnetic systems.
 
 \end{abstract}

\maketitle

Custom-designed artificial magnetic materials provide model platforms for studying a variety of fundamental problems in magnetism and also form the basis for applications such as non-volatile memory and spin-based logic. Examples include magnetic multilayers and heterostructures wherein interlayer and interfacial exchange coupling yields new magnetic behavior relevant to spintronic applications \cite{Hellman2017}, continuous ferromagnetic films interfaced with patterned ferromagnetic arrays that allow systematic control over pinning of magnetic domain walls \cite{Metaxas2009}, and nanoscale ferromagnetic elements with mixed anisotropy that allow the engineering of spin frustration and spin texture \cite{Luo2019}. Artificial spin ice (ASI) is another interesting example of a custom-designed magnetic material, in which interacting arrays of lithographically-defined single domain nanomagnets are arranged in frustrated two-dimensional geometries (e.g. triangular and kagome) \cite{Wang2006, Skjarvo2020}.  A variety of fundamental phenomena have been studied in ASIs, including the physics of Dirac strings and magnetic monopoles \citep{Ladak2011, Mengotti2011,Phatak2011,Mol2009,Perrin2016}, the interplay between frustration and thermal fluctuations \citep{Kapaklis2014,Farhan2013}, and geometric \cite{Nisoli2013,Heyderman2013a}, and topological \cite{Drisko2017} and vertex \cite{Gilbert2014a} frustration. ASI systems are also candidates for reconfigurable magnonics and neuromorphic computing. 

Perpendicular ansitropy ASIs, in which the individual islands have perpendicular magnetic anisotropy (PMA) with moments pointing normal to the plane of the structure, \cite{Mengotti2009, Zhang2012a} are of particular interest since the entire microstate can be probed in full detail using spatially-resolved magneto-optical Kerr effect (MOKE) microscopy \cite{Fraleigh2017,Kempinger2020}. This is because the polar MOKE effect has a large enough magnitude to readily allow extreme diffraction-limited microscopy with the spatial resolution ($\sim 300 $ nm) required to measure the magnetic state of each individual island in an array. However, perpendicular anisotropy ASIs have weaker inter-island coupling than their in-plane counterparts, making it difficult to access low-energy states. In this Letter, we show how the effective geometry can be manipulated without changing the physical geometry by altering the interactions between islands. We do this by fabricating a composite system comprised of an array of ferromagnetic islands with PMA supported on a 15 nm layer of the soft ferromagnet Ni$_{0.80}$Fe$_{0.20}$, also known as permalloy (Py). In bit patterned media, such underlayers have been shown to increase the interaction strength of arrays \cite{Litvinov2000,Litvinov2001,Ek2001}. Micromagnetic modeling has been useed to show that for an in-plane artificial spin ice array, coupling to a soft underlayer leads to a complex system exhibiting dynamically coupled modes \cite{Iacocca2020}. Because the underlayers must be soft enough to be affected by the stray field of the magnetic islands, they can be biased by even a modest in-plane field. This breaks the lateral symmetry of the system, resulting in new ground state properties, including
the ability to select \textit{in situ} a subset of the microstates that the array can achieve during demagnetization. The underlayer also tunes the relative coupling strength of the nearest neighbors (NN) and next nearest neighbors (NNNs). For arrays with a square geometry, this can lead to an effective NN coordination of 8, a geometrical impossibility in a two dimensional system. Our results demonstrate that soft ferromagnetic underlayers can be used to build highly tunable systems, both enabling  model systems with properties that have not been experimentally realized otherwise and also opening the door to possible new applications.

\begin{figure} [ht]
\centering
\includegraphics[width = .8\columnwidth]{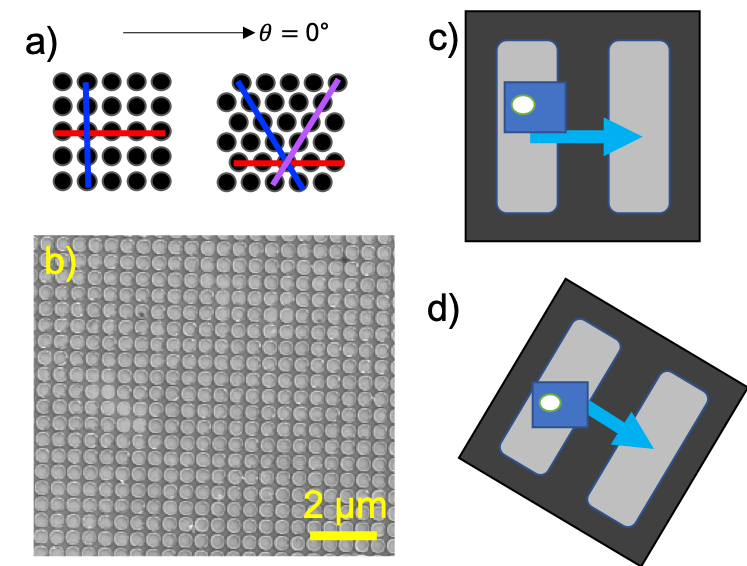}
\caption{a) Square and Triangular geometries with lines showing the principal axes of the lattices. b) Scanning electron microscope image of perpendicular magnetic anisotropy islands on Py in a square lattice array. c) Schematic of sample over projected field magnet. The white dot represents the illuminated area for the MOKE studies, the grey bars represent the magnet poles, and the blue arrow represents the direction of the in-plane component of the magnetic field. d) Schematic with the magnet rotated to a different position, demonstrating how the in-plane field is rotated in direction.}
\label{Experimental}
\end{figure}

We fabricated two separate samples of Pt/Co multilayer islands on Py underlayers, which were demagnetized and measured independently. 15 nm thick Py underlayers were first deposited using electron beam evaporation, and then Ti(2 nm)/Pt(10 nm)/[Co(0.3 nm)/Pt(1 nm)]$_8$ islands were deposited using electron beam lithography and DC sputtering. The islands were 425 nm in diameter, with lattice spacing ranging from 500 nm to 800 nm, in both frustrated (triangular) and unfrustrated (square) geometries. The triangular lattice had sixfold rotational symmetry, and the square lattice had fourfold rotational symmetry (See Fig. \ref{Experimental}a). We also studied kagome (frustrated) and hexagonal (unfrustrated) geometries, both of which are decimations of a triangular lattice. Data for these arrays are included in Supplemental Materials \footnote{See Supplemental Material at URL for additional simulated and experimental data.}. SEM imaging confirmed the size of the islands and showed that neighboring islands were well defined and separated (Fig. \ref{Experimental}b). 

To probe these systems, we used a custom-built diffraction-limited, polar MOKE microscope with a resolution of 300 nm to collect images of demagnetized states of arrays for different lattice types and spacings. The external field was controlled using a 2-pole projected field magnet (GMW5201) with a dominant perpendicular field component and a smaller in-plane component. We prepared the states with an oscillating, decreasing magnetic field protocol with a maximum applied field of 1.5 kOe and a step size of 2 Oe in the switching region, using in-plane field angles between $-60^{\circ}$ and $60^{\circ}$. Because the PMA structures are not robust against heating, this sort of demagnetization protocol is standard for preparing perpendicular anisotropy ASI in a low energy collective state \cite{Zhang2012a}.

The coercive field of the array was around 800 Oe. The in-plane field is approximately 5\% of the out-of-plane field and the coercivity of the soft underlayer is very small ($<$ 10 Oe). As a result, throughout the switching region, the in-plane field component is sufficient to saturate the soft underlayer and the magnetization direction of the underlayer is reversed with each reversal of the perpendicular field direction. The direction of the in-plane field component is controlled by rotating the magnet underneath the sample, as shown in Fig. \ref{Experimental}c,d. Measurement angles are chosen such that the sample stays over the magnet pole during rotation; angles that are too large cause the sample to shift away from the pole and no longer experience a strong perpendicular field. In the extended arrays, we define $\theta$ to be the angle between the in-plane field component and  one of the principal lattice directions, as shown in Fig. \ref{Experimental}.  Micromagnetic simulations demonstrated that the underlayer  significantly enhances the interaction strength, as discussed in Supplemental Materials \cite{Note1}.

\begin{figure} [ht]
\centering
\includegraphics[width = .8\columnwidth]{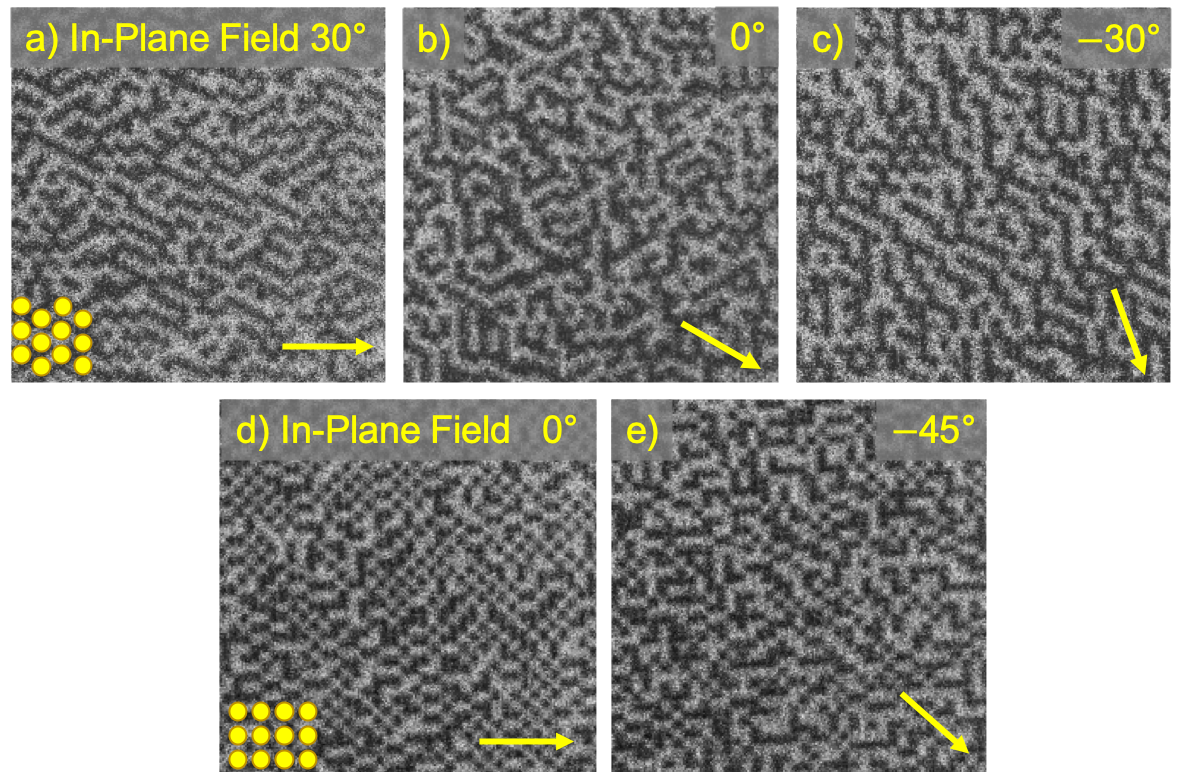}
\caption{MOKE images of perpendicular magnetic anisotropy artificial spin ice after demagnetization for triangular and square lattice geometries.   Top row:Images of demagnetized 500 nm triangular array with in-plane field oriented at a) $\theta$ = 30$^{\circ}$, b) $\theta$ = 0$^{\circ}$, and c) $\theta$ = -30$^{\circ}$.  Bottom row: MOKE images of demagnetized 500 nm square array with in-plane field oriented at d) $\theta$ = 0$^{\circ}$ and e) $\theta$ = -45$^{\circ}$.  The dots in panel (a) and panel (d) represent the orientation of the lattices within the images. }
\label{MokeImages}
\end{figure}

Fig. \ref{MokeImages} shows MOKE images of a triangular (a, b, c) and a square (d, e) array demagnetized at various in-plane field orientations. To generate images with strong magnetic contrast, we subtract images of the array in a saturated state from images in the demagnetized state. All images are taken in the remanent state. After the background subtraction, islands that are ``up” and ``down” appear light and dark respectively. The Py magnetization is not visible, as it lies in the plane. The triangular lattice as shown in the MOKE images is rotated 30$^{\circ}$ from the orientation shown in Fig. \ref{Experimental} (see Fig. \ref{MokeImages}). 
In the triangular array, the variation in nearest neighbor correlation is observable by eye in images of the different microstates. Fig. \ref{MokeImages} a,b,c show the demagnetized state of a 500 nm spacing triangular array prepared with $\theta =$ 30$^{\circ}$, 0$^{\circ}$, and $-30^{\circ}$ respectively. We observe that the microstates corresponding to 30$^{\circ}$ and $-30^{\circ}$ are qualitatively similar, with extended lines of aligned islands in the direction of the applied field. At $\theta =$ 0$^{\circ}$, however, the microstate appears more disordered. In the square array, the data with $\theta =$ 0$^{\circ}$ shows large patches of the ground state checkerboard pattern as expected. However, the $-45^{\circ}$ data appears quite different, with extended regions of aligned islands in the direction of the applied magnetic field, similar to the ordered state observed in the triangular array. It is visually apparent from the raw data that there is a significant impact on the microstate when the direction of the in-plane field is changed.

To characterize the order in these states, we calculate the NN correlation function
\begin{equation}
C_S = \langle S \rangle ^2 - \sum_{i,j} {S_i} {S_j}
\label{correlations}
\end{equation}

The sum is taken over NN pairs, and $S$ is the magnetization direction of each island (up or down, denoted as $\pm$1). This form matches the form used in previous publications \cite{Kempinger2020}. $\langle S \rangle ^2$ is nearly zero in demagnetized states, hence provides only a minor correction. 

\begin{figure} [ht]
\centering
\includegraphics[width=\columnwidth]{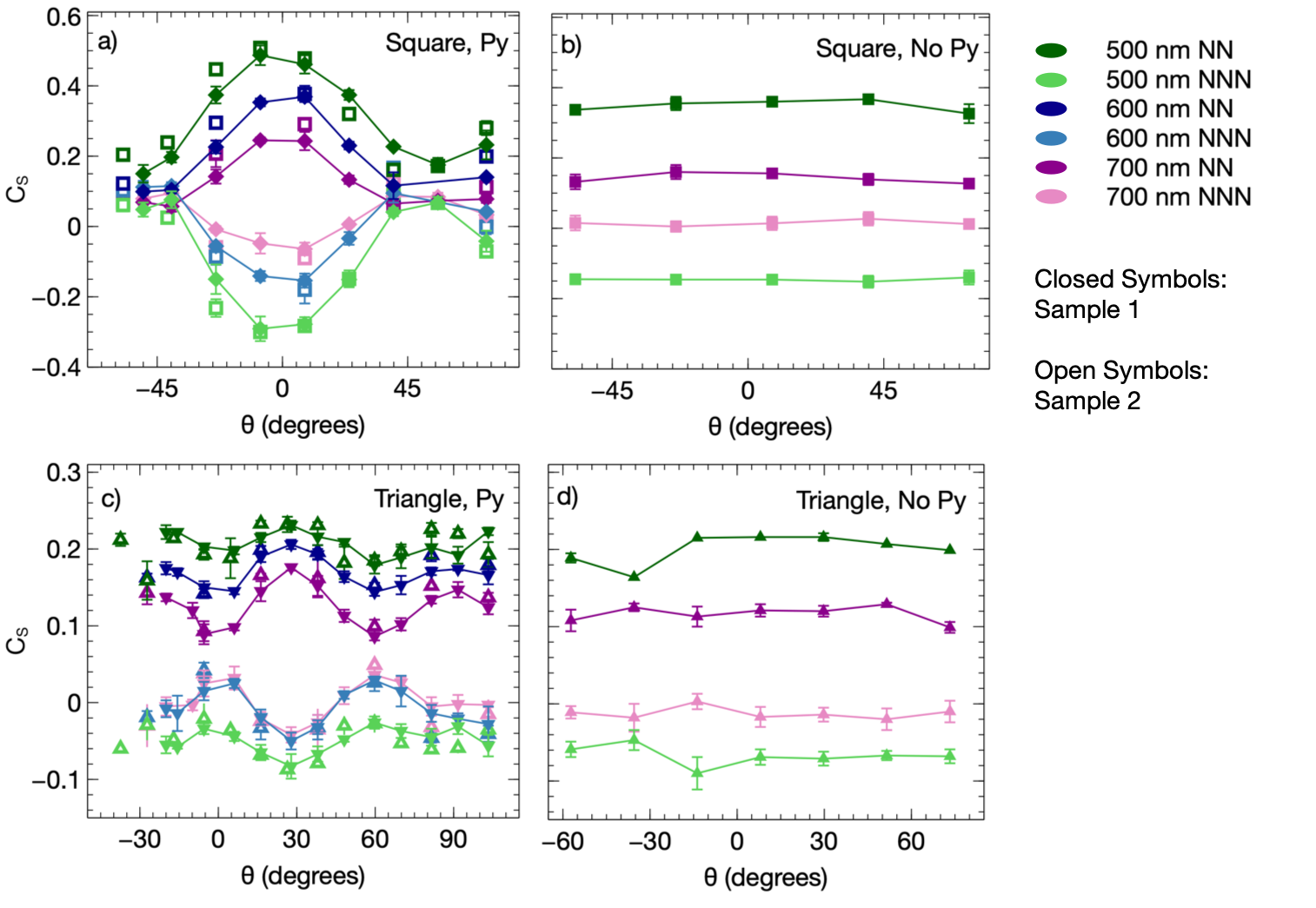}
\caption {Moment correlations of square and triangular arrays as a function of the direction of the in-plane component of the magnetic field during demagnetization, with and without permalloy underlayers.}
\label{FullMicrostate}
\end{figure}

The correlation values for samples with Py underlayers show a clear variation of $C_S$ with in-plane field angle for all lattices considered (Fig. \ref{FullMicrostate}a,b). The control samples without Py underlayers  show no such variation (Fig. \ref{FullMicrostate}c,d). The directional dependence is clearly an effect of the soft underlayer. 

For the triangular array, the correlation is maximized at $-30^{\circ}$, 30$^{\circ}$, and 90$^{\circ}$, i.e., when the field is normal to a principal lattice direction. These correspond to the states with extended lines of aligned islands shown in Fig. \ref{MokeImages}. 
Similar behavior is seen in the other sixfold symmetric lattices discussed in Supplemental Material \cite{Note1}. At 0$^{\circ}$ and 60$^{\circ}$, the field is oriented along one of principal lattice directions and normal to one of the \textit{next} NN directions. These angles correspond to the disordered triangular state in Fig. \ref{MokeImages}. At these angles there is a decrease in NN correlation and a corresponding increase in next nearest neighbor (NNN) correlation. This variation becomes more pronounced at larger lattice spacings. This indicates that there might be a significant influence of NNN coupling on the demagnetized states of the array. 

The correlation in the square array has a much larger $\theta$ dependence, which is accompanied by a dramatic variation in the NNN correlation. In the ground state, the NNN correlation is ferromagnetic, mediated by the antiferromagnetic NN correlation. However, when the field is aligned at $\pm 45^{\circ}$, normal to an NNN direction, the NNN coupling becomes antiferromagnetic.

To understand if this effect is consistent with expectations, we turn to micromagnetic simulations. We used the micromagnetics package MuMax3\cite{Vansteenkiste2014} to simulate a single pair of islands with and without a Py underlayer. Simulation data and details can be found in Supplemental Material \cite{Note1}. When carrying out the simulations with underlayers, the interaction energy of a pair of islands depended on the angle between the in-plane field and the pair of islands. To extrapolate to an extended line of islands, we used appropriate symmetry considerations along with an assumption that a demagnetized line of islands contains an approximately even mixture of ``up" and ``down" islands. These considerations are discussed in detail in Supplemental Materials \cite{Note1}. 
There is a large peak in interaction energy when the field is oriented perpendicular to the pair of islands in agreement with our expectations. When the field is oriented parallel to the pair of islands, there is a significant dependence of the energy on whether the field points from the ``up'' island to the ``down'' island or vice versa . This difference leads to a second smaller peak in overall interaction energy when the data is extended to a line. 
This suggests that the correlation in an extended line of islands can be described by 

\begin{equation}
C_{S, \textrm{directional}} = a\cos(2\phi)+b\cos(4\phi)+c
\label{coseqn}
\end{equation}

where $\phi$ = 90$^{\circ}$ represents the external field aligned along the line of islands.

%
%
%
%

\begin{figure*} [ht]
\centering
\includegraphics[width = .9\textwidth]{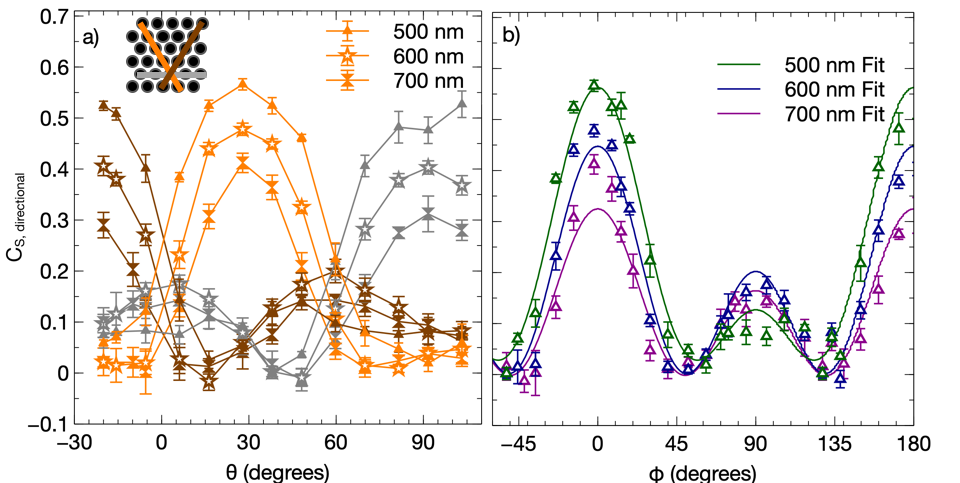}
\caption{a) Nearest neighbor moment correlations in the triangular lattice calculated considering nearest neighbor pairs oriented along the principal axes D1, D2, and D3 as a function of $\theta$ for three different lattice spacings. b) The same data combined by symmetry arguments into a single curve for each lattice spacing, showing how the correlation in a given direction varies with the angle of the in-plane field component. Lines in (b) are lines of best fit to Eqn. \ref{coseqn}.}
\label{DirectionalCorrTri}
\end{figure*}

\begin{figure*} [ht]
\centering
\includegraphics[width = .9\textwidth]{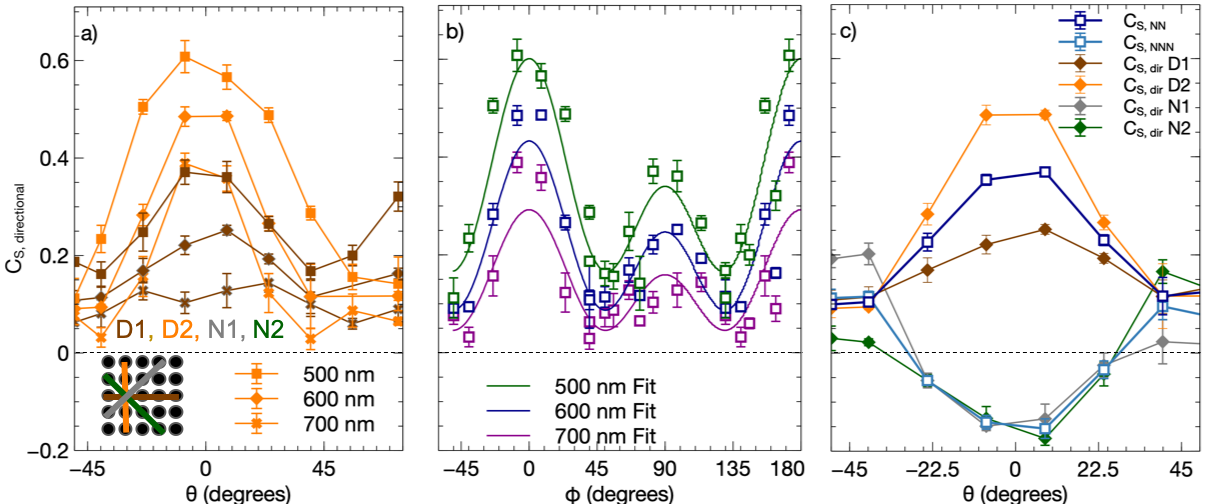}
\caption{a) Nearest neighbor moment correlations in the square lattice considering nearest neighbor pairs oriented along principal axes D1 and D2 as a function of $\theta$ for three different lattice spacings. b) The same data combined by symmetry arguments into a single curve for each lattice spacing, showing how the correlation in a given direction varies with the angle of the in-plane field component. Lines in right graphic are lines of best fit to Eqn. \ref{coseqn}. c) Nearest neighbor and next-nearest-neighbor correlation data for each lattice direction, along with overall nearest neighbor and next-nearest-neighbor correlations for the 600 nm square array.}
\label{DirectionalCorrSqu}
\end{figure*}

Eqn. \ref{coseqn} gives us the information we need to additionally consider directional dependence of correlations. We considered lines of islands in D1, D2, and D3 for the triangular lattice (as defined in Fig. \ref{DirectionalCorrTri}a, subset) and D1 and D2 for the square lattice (as defined in Fig. \ref{DirectionalCorrSqu}a, subset). The data for each of these groups is combined into a single curve by considering the angle between the applied field and the line under consideration, with $\phi$ = 90$^{\circ}$ representing the external field aligned along the line. This combined data is fit to Eqn. \ref{coseqn}
(See Fig. \ref{DirectionalCorrTri}b and \ref{DirectionalCorrSqu}b, respectively). Best fit parameters and $\chi^2$ values can be found in Supplemental Material \cite{Note1}.

The directional correlation values for the triangular lattice match our visual observations of ordered states occurring at $\theta = -30^{\circ}, 30^{\circ}$. The average correlation ($C_S$) of a moment array with a frustrated geometry cannot exceed 1/3, because the competing interactions prevent perfect ordering \cite{Zhang2012a}. Without the soft underlayer, this average correlation is the same for  all three NN directions (see Supplemental Material \cite{Note1}). With the symmetry-breaking underlayer however, the maximum value of directional correlation ($C_{S,\textrm{directional}}$) in the triangular lattice reaches values as large as 0.55, as shown in Fig. \ref{DirectionalCorrTri}a. 
This increase of $C_{S,\textrm{directional}}$ in one direction leads to a suppression of $C_{S,\textrm{directional}}$ in other directions so that the average value of $C_S$ obeys the appropriate limit. A high value of correlation occurs when there is strong antiferromagnetic ordering, which occurs in the direction perpendicular to the applied field, leading to the apparent chains of aligned islands in the applied field direction for angles with highly ordered states. Considering $C_{S,\textrm{directional}}$ as a function of interaction strength, we observe that the height of the main peak (0$^{\circ}$ in Fig. \ref{DirectionalCorrTri}b) decreases steadily with decreasing interaction strength, while the minimum value consistently shows almost no correlation and the secondary peak (90$^{\circ}$ in Fig. \ref{DirectionalCorrTri}b) maintains approximately the same height. The kagome lattice data shows similar behavior (see Supplementary Material \cite{Note1}). In contrast, the square array (Fig. \ref{DirectionalCorrSqu}b) and hexagonal array (see Supplementary Material \cite{Note1}) show large, decreasing peaks, but also show an overall decrease in the minimum correlation and a suppression of the secondary peak with decreasing interaction strength. 

The square lattice differs from all the other lattices considered in having fourfold, rather than sixfold, rotational symmetry. As shown in Fig. \ref{FullMicrostate}, both the NN and NNN correlations in this array show a pronounced variation with applied field angle. 
The variation in NNN correlation is strong enough to show a noticeable splitting between different NNN directions (N1 and N2, Fig. \ref{DirectionalCorrSqu}c). For the 500 nm lattice spacing shown in the images in Fig. \ref{MokeImages}, the NN and NNN correlation are not significantly different at points of minimum nearest-neighbor correlation. As the lattice spacing increases, we begin to observe points where the next-nearest-neighbor correlation exceeds the nearest neighbor correlation at the minimum values of NN correlation. If we take the correlation to be reflective of the effective interaction strength, this indicates the existence of a crossover point where the NN and NNN interactions  are equally influential. At the crossover point there is an induced frustration in the array, because the system no longer has a preference between NN and NNN correlations. This explains why the square array demagnetized with $\theta$ = 45$^{\circ}$ looks qualitatively similar to demagnetized states of the triangular array. 45$^{\circ}$ is the angle at which NN and NNN correlations differ least in the 500 nm lattice. Since the NN correlations are no longer dominant, the pattern of the demagnetized array is no longer best characterized as regions of the ground state. Instead, the induced frustration allows it to develop the same characteristics as the frustrated triangular array. The angle of the extended lines of islands is the visual manifestation of the splitting in NNN correlation.


The convergence (or even inversion) of the NN and NNN correlation values has some strange implications. The correlation values arise from the competition between interaction strength and disorder in the array. Disorder in artificial spin ice with perpendicular anisotropy is primarily a single-island property \cite{Fraleigh2017}, which should not have a directional dependence. Therefore, within a single fabricated sample, the character of the variation in correlation as the in-plane field is rotated should be governed primarily by the island-island interaction. Our micromagnetic simulations also support this conclusion. From this, we conclude that as the NN and NNN correlation values approach one another, the effective NN and NNN interactions, as modified by the underlayer, produce a system in which the distinction between NN and NNN is blurred, and insofar as correlation is concerned, effectively lost: at the crossover point, the square lattice behaves as if it had an effective coordination number of 8. Close-packed two-dimensional systems have a coordination number of 6, so a coordination number of 8 is theoretically not possible for a two dimensional system. 

Our results demonstrate that the addition of a soft magnetic underlayer to a perpendicular anisotropy ASI system enhances the interaction strength and breaks the rotational symmetry of the array by adding a directional dependence to the NN coupling and enhancing the NNN coupling. These two effects together cause the correlation in the array to vary with the angle of an in-plane field applied during demagnetization, providing a new mechanism through which to control the behavior of ASI systems, specifically the selection of a subset of available microstates during a demagnetization process. Both frustrated and unfrustrated lattices demonstrate a directional dependence of the correlation influenced by NNN correlations, and the frustrated lattices (triangular and kagome) show an increased impact of NNN effects as the lattice spacing is increased. This phenomena will be further investigated in future studies. The observation of an effective coordination number of 8 in a two-dimensional system indicates that this additional control  will open the already-flexible ASI systems into a new dimension of parameter space that will allow the testing of spin models that would not otherwise be experimentally accessible. Furthermore, the reconfigurable components of ASI with this additional degree of control could enhance the new device concepts that are being considered in neuromorphic computing and magnonics. 

This project was funded by the US Department of Energy, Office of Basic Energy Sciences, Materials Sciences and Engineering Division under Grant No. DE-SC0010778 and DE-SC0020162.Sample fabrication at Argonne National Laboratory was supported by the US Department of Energy, Office of Science, Basic Energy Science, Materials Science and Engineering Division. The use of the Center for Nanoscale Materials was supported by the U.S. Department of Energy (DOE), Office of Science, Basic Energy Science (BES), under Contract No. DE-AC02-06CH11357. We acknowledge DMR-1420620 for support of analysis and modeling efforts

\bibliography{Kempinger_2021}
\bibliographystyle{apsrev4-1.bst}

\newpage
\beginsupplement
\section*{SUPPLEMENTARY INFORMATION for ``Field-tunable interactions and frustration in underlayer-mediated artificial spin ice" by S. Kempinger {\textit et al}}

\section{SQUID magnetometry}

\begin{figure} [h!]
\centering
\includegraphics[width = .5\textwidth]{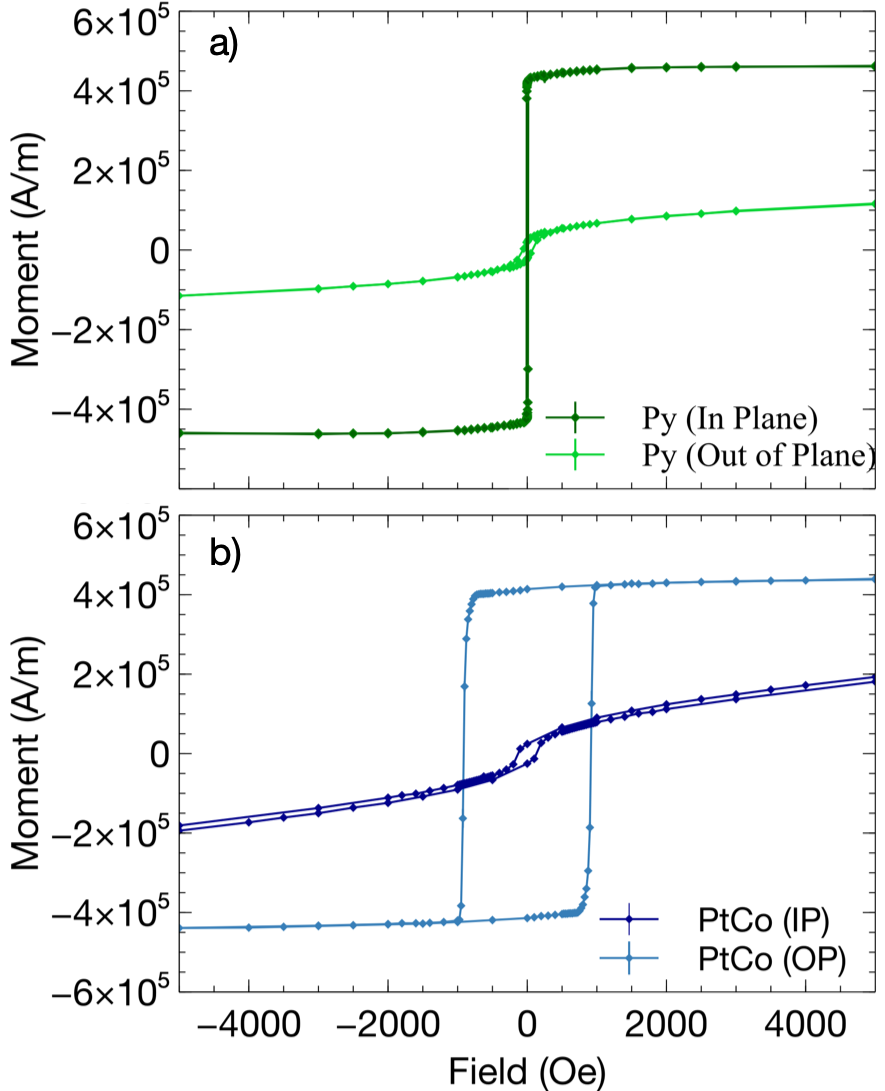}
\caption{Experimental SQUID magnetometry data.  a) M vs. H response of a continuous Py film measured at 300 K with the magnetic field oriented both parallel (dark green) and perpendicular (light green) to the film surface. b) M vs. H response of a continuous PtCo film measured at 300 K with the magnetic field oriented both parallel (dark blue) and perpendicular (light blue) to the film surface.}
\label{SQUID}
\end{figure}

Measurements were taken using a Quantum Design MPMS system with the external field applied both parallel and perpendicular to the film plane. The PtCo film is a continuous film co-deposited with one of the patterned samples discussed in the text. The Py film is a continuous film co-deposited with one of the films that was used as an underlayer. Samples were measured at 300 K. PtCo samples were measured over a field range of $\pm$2 T, with fine steps of 25 Oe in the switching region around 800 Oe. Py samples were measured over a field range of $\pm$ 1 T, with fine steps of 2 Oe in the switching region around 0 Oe. PtCo samples show strong perpendicular anisotropy, with square hysteresis and a coercivity of approximately 800 Oe. Py samples show in-plane anisotropy with low coercivity ($<$10 Oe). The hysteresis visible in the out-of-plane measurement of the Py sample is due to magnetic contamination in the Si substrate.

\newpage

\section{Micromagnetics}

\begin{figure} [ht]
\centering
\includegraphics[width = .7\textwidth]{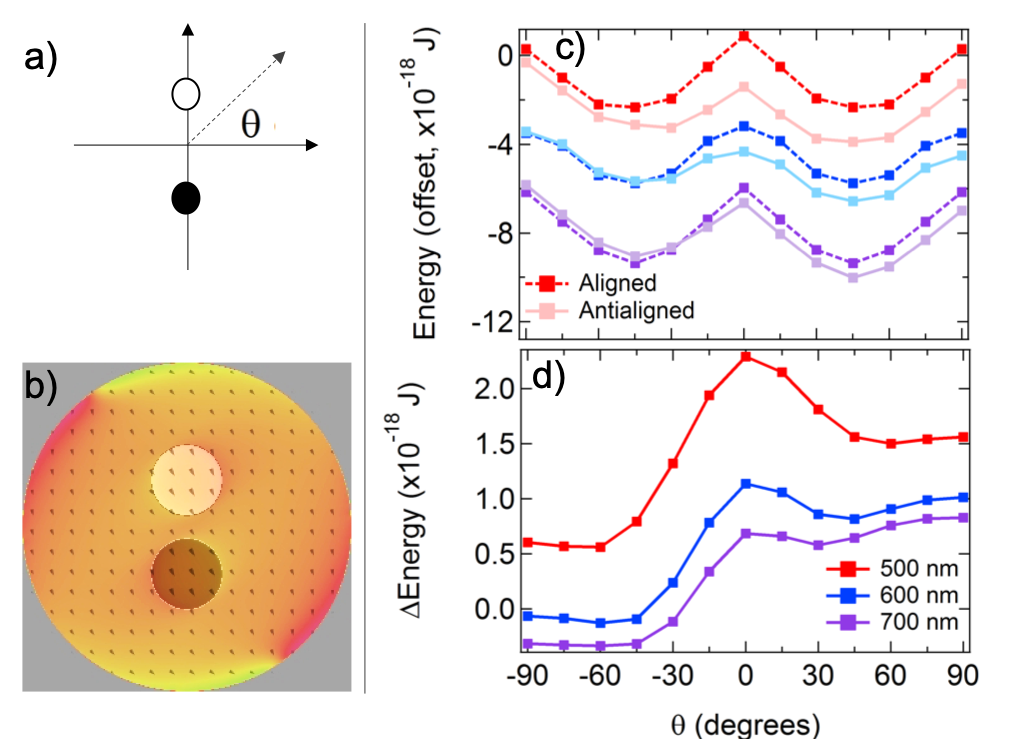}
\caption{Micromagnetic data on sample with Py underlayer. a) Schematic showing the definition of the angle $\theta$ used in micromagnetic simulations. b) Micromagnetic image of a pair of magnetically anti-aligned islands over a permalloy underlayer magnetized at $-45^{\circ}$ after relaxation. c) Energy in aligned and anti-aligned pairs of islands after relaxation for 500 nm, 600 nm, and 700 nm spacing, offset so data is not overlapping. d) Energy difference between aligned and anti-aligned pairs of islands, i.e. interaction energy. }
\label{Micromagnetics}
\end{figure}

We use the micromagnetics package MuMax3\cite{Vansteenkiste2014}  to explore the effect of adding a soft magnetic underlayer on the energy landscape of an isolated pair of perpendicular nano-islands. We simulate one pair of 450 nm diameter islands ($M_S$ = 3.5 e 5 A/m, $K_1$ = 94 e 3 J/$m^3$, $A$ =1 e -11 J/m,) on a 2 $\mu$m diameter, 15 nm thick layer of Py ($M_S$ = 8.0 e 5 A/m, $A$ = 1.3 e -11 J/m). We use a round shape for the Py underlayers to minimize edge effects. While real islands consist of many thin layers, simulations treat the island as one effective magnet with a thickness of 10.4 nm, and a 12 nm gap between the island and the Py to represent the non-magnetic buffer layer. Island parameters were determined by SQUID measurements of Pt/Co films. Island to island spacings are simulated over a range from 500 nm to 800 nm. We initialized islands in the pair with either the same magnetization or opposite magnetization, and initialized the Py with a random magnetization. We allowed the magnetization to relax in an external field large enough to saturate the Py, as our experimental data indicates the in-plane field component is large enough to saturate the Py layer throughout the switching region. An example of such a relaxed state is shown in Fig. \ref{Micromagnetics}b. Final energy values are extracted from both aligned and antialigned states (see Fig. \ref{Micromagnetics}c), and interaction energy is calculated as the difference between these values ($\Delta$Energy, Fig. \ref{Micromagnetics}d). A value of $\Delta$Energy $>0$ indicates that the energy is higher in the aligned state, and $\Delta$Energy $<0$ indicates the energy is higher in the antialigned state. So positive values of $\Delta$Energy correspond to preferred antiferromagnetic ordering, which is the expectation for an isolated pair of interacting islands.

In agreement with our predictions, the addition of a Py underlayer during a demagnetization process in the presence of an external in-plane field  increases the interaction strength within the pair of islands as well as breaking the lateral symmetry of the system. The angle of the field in these simulations is defined to be $0^{\circ}$ when the in-plane field is perpendicular to the pair of islands as shown in Fig. \ref{Micromagnetics}a. In a pair of islands with 500 nm spacing, there is in general an energetic preference for the antiferromagnetic state over the ferromagnetic state, demonstrated by the positive values of $\Delta$Energy in Fig. \ref{Micromagnetics}d. The largest increase in interaction strength occurs when the external magnetic field is normal to the island pair, at $0^{\circ}$. Surprisingly, however, as the island spacing increases, the energy difference between aligned and antialigned states is close to zero as shown in the 600 nm spacing from $-90^{\circ}$ to around $-45^{\circ}$ or even inverted as shown in the 700 nm spacing in the same range (see Fig. \ref{Micromagnetics}d).  This indicates that, in such a configuration, the pair of islands prefers the ferromagnetic state over the antiferromagnetic state. These simulations suggest that we can change both the coupling strength and break the lateral symmetry in perpendicular artificial spin ice, but also that we can selectively tune the type of coupling between neighbors from antiferromagnetic to ferromagnetic (See Fig. \ref{Micromagnetics}). 

In the micromagnetic simulations, we considered only a single pair of islands with the magnetic field at $0^{\circ}$ oriented from ``up" to ``down". To generalize these observations to effects in a lattice, we need to consider the impact of geometry. In an extended, demagnetized line of islands, there will be an even mixture of ``up" and ``down" islands, and thus a given magnetic field direction will be oriented at $\theta^{\circ}$ to some pairs and $(180+\theta)^{\circ}$ to others, and these two situations should occur with approximately the same prevalence. 

Additionally, a single pair of islands has mirror symmetry. So the interaction energy of a pair of islands with an in-plane field oriented at $\theta^{\circ}$ is the same as the interaction energy of a pair with field oriented at $(180-\theta)^{\circ}$ and, also, the interaction energy with the field oriented at $(180+\theta)^{\circ}$  is the same as at $-\theta^{\circ}$.
This suggests we can construct the expected behavior of an extended line of islands by averaging together the micromagnetic data at $\theta^{\circ}$ and $-\theta^{\circ}$ to estimate the total energy in the line, which we assume to be proportional to the correlation in the line. This averaging gives a large peak in interaction energy at $0^{\circ}$ and a smaller secondary peak at $-90^{\circ}$ and $90^{\circ}$.
The symmetry requires a Fourier expansion including only even multiples of $\theta$, and fitting the data from micromagnetic simulations requires at least two non-zero terms, so we use the following function to describe the correlation in an extended line of islands:

\begin{equation}
C_{S, \textrm{directional}} = a\textrm{cos}(2\theta)+b\textrm{cos}(4\theta)+c
\label{coseqn_SI}
\end{equation}

\newpage
\section{Estimate of increased interaction strength from Py underlayer}

To consider the increase in interaction strength of a pair of islands, we simulate the pair of islands with and without an underlayer. To compare to the dipole approximation, we convert the interaction energy to the effective field in G of one island in the pair at the center of the other island. The interaction strength without an underlayer in simulations agrees relatively well with the dipole approximation, although the dipole approximation underestimates the interaction strength at small lattice spacings. The addition of an underlayer increases the interaction strength in the simulations by approximately a factor of 2.

\begin{figure} [ht]
\centering
\includegraphics[width = .7\textwidth]{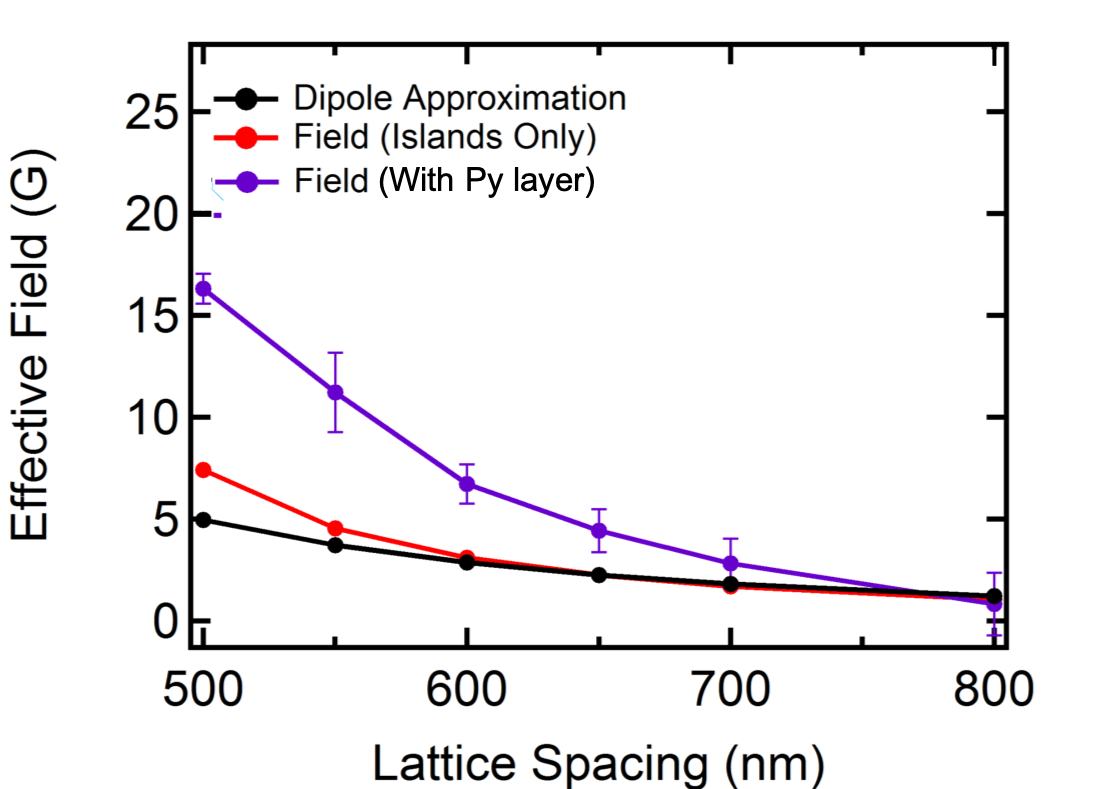}
\caption{Interaction strength converted to the effective field of an island on it's neighbor using the dipole approximation and simulations with and without a Py underlayer.}
\label{InteractionStrength}
\end{figure}

To verify the increase experimentally, we consider the correlation in the demagnetized states with the external field aligned such that the correlation is at its maximum value. These are measured on two different samples, each with a corresponding control sample that was simultaneously fabricated without a Py underlayer. Measurements shown here are on square and hexagonal lattices with spacings ranging from 500 nm to 800 nm. Using methodology developed elsewhere \cite{Kempinger2020}, we construct a dimensionless x-axis using the interaction strength from the dipole approximation and the measured disorder in the arrays. The disorder is measured from the width of the switching distributions during hysteresis measurements. Using just the dipole field of the islands in the construction of the x-axis, the samples with a Py underlayer show an enhancement of correlation over the samples without. Using the increase in interaction strength by a factor of 2 suggested by the micromagnetic simulations, the data for samples with and without underlayers follow the same trend. This experimentally supports the data from simulations that the underlayers increase the interaction strength by approximately a factor of 2. 

\begin{figure} [ht]
\centering
\includegraphics[width = .7\textwidth]{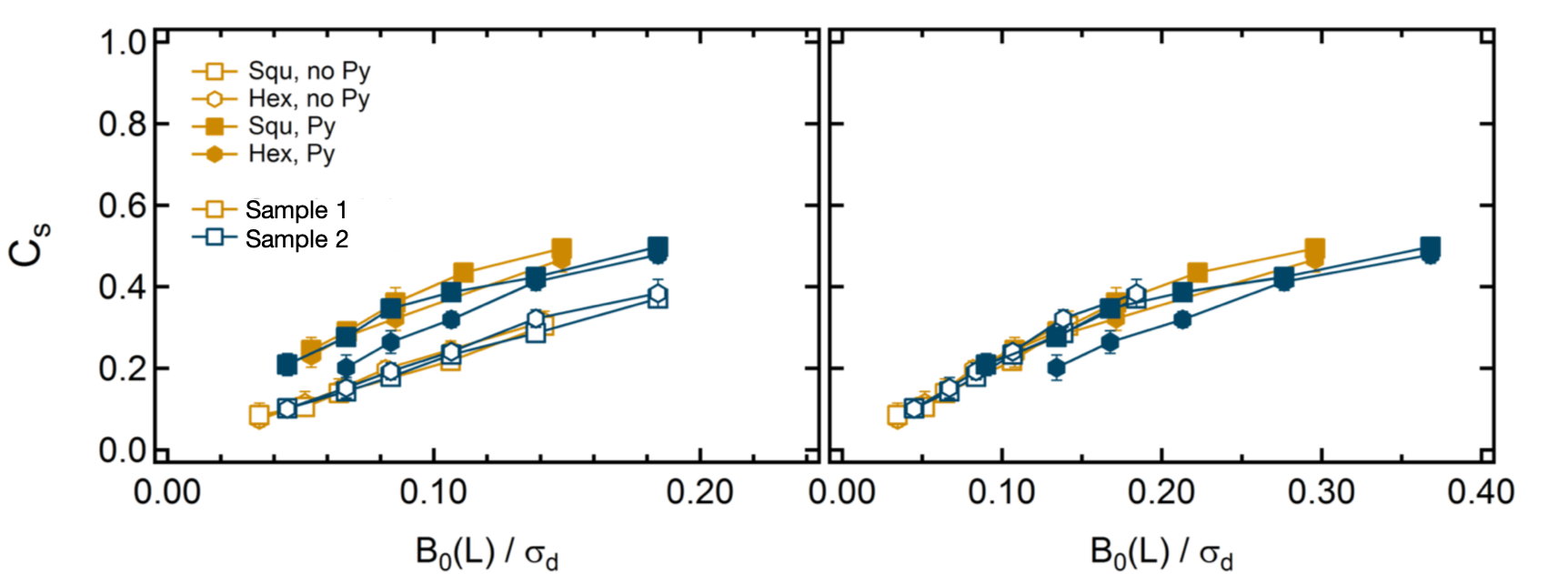}
\caption{Left panel: correlation as a function of interaction strength scaled by disorder using the dipole approximation to calculate the interaction strength. Right panel: correlation as a function of interaction strength scaled by disorder, assuming that the interaction strength for samples with Py underlayers is approximately 2x the interaction strength given by the dipole approximation.}
\label{InteractionStrengthCorrelation}
\end{figure}

\newpage
\section{Samples without soft underlayers}

\begin{figure} [ht]
\centering
\includegraphics[width = .7\textwidth]{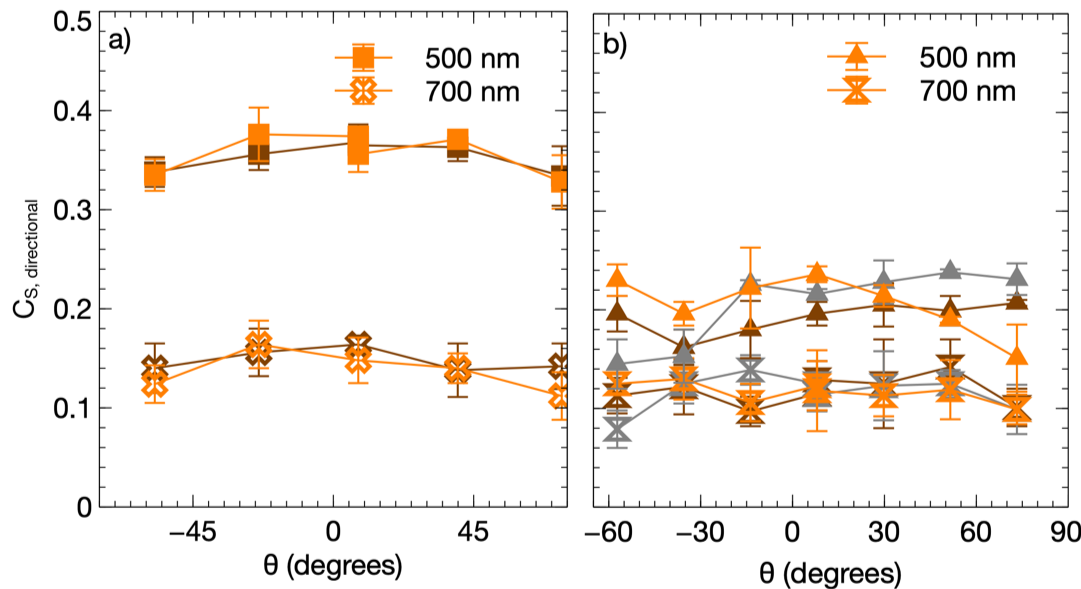}
\caption{a) Directional correlations for 500 nm and 700 nm lattice spacing square geometry arrays in a sample without a Py underlayer. b) Directional correlations for 500 nm and 700 nm lattice spacing triangular geometry arrays in a sample without a Py underlayer. The colors correspond to different directions using the same color scheme as Fig. \ref{DirectionalCorrTri} and \ref{DirectionalCorrSqu} in the text.}
\label{DirectionalCorrControl}
\end{figure}

The directional dependence of correlations in the control samples is not split between different lattice directions, as can be see in Fig. \ref{DirectionalCorrControl}.  This is expected  because there is no symmetry-breaking underlayer to introduce a preferred direction of correlation. The value of correlation in each direction agrees with the overall value of correlation measured for the whole array (Fig. \ref{FullMicrostate} in the text). There is more variation between directions in the triangular array since the state is less ordered due to the frustration in the geometry. The variation is not systematic and does not indicate any preferential direction for the correlations. The directional correlation in the triangular array from the control sample never exceeds the average value of 1/3 for a frustrated array.

\newpage
\section{Hexagonal Lattice Data}

The hexagonal lattice is a decimated version of a sixfold rotationally symmetric lattice, in an unfrustrated geometry. It shows a similar level of variation in the correlations for all lattice spacings considered. Because it is not a frustrated geometry, it does not appear to have increasing effects of next nearest neighbor correlation at larger lattice spacings. 

\begin{figure} [ht]
\centering
\includegraphics[width = .7\textwidth]{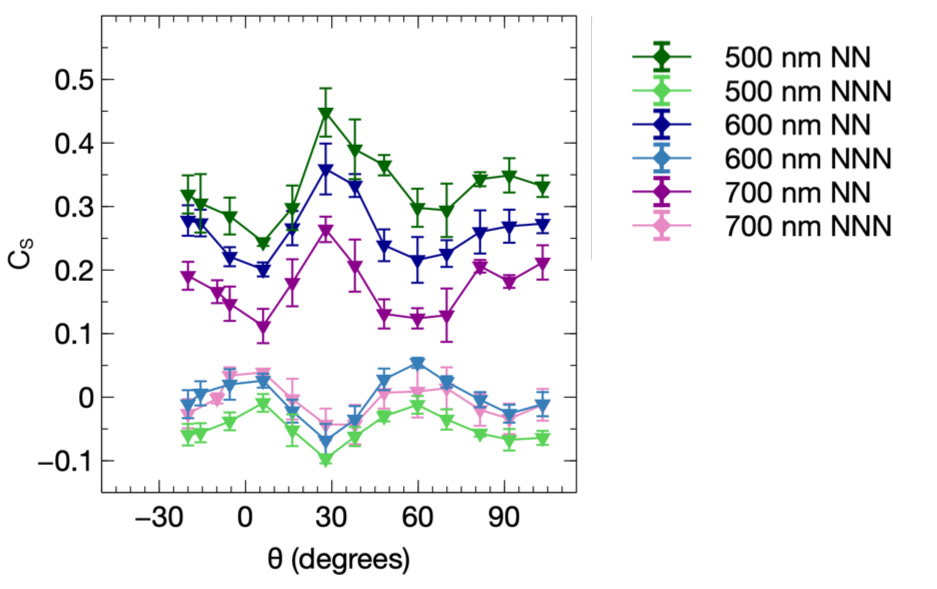}
\caption{$C_S$ values for hexagonal lattice with three different lattice spacings}
\label{Hexagonal}
\end{figure}

\begin{figure} [ht]
\centering
\includegraphics[width = .7\textwidth]{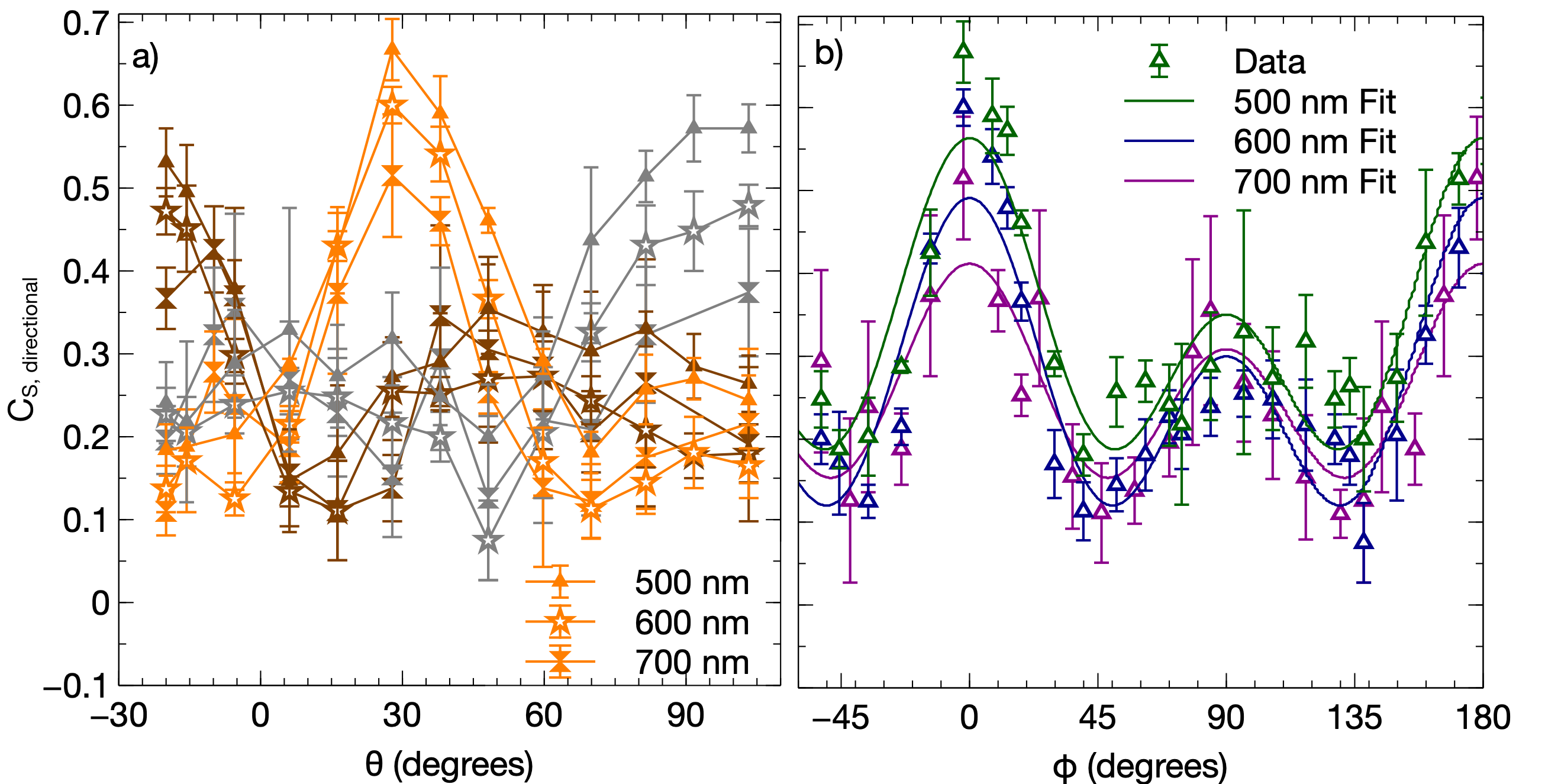}
\caption{a) Nearest neighbor correlation in hexagonal lattice calculated considering NN pairs oriented along D1, D2, and D3 as defined in the text as a function of $\theta$ for three different lattice spacings. b) The same data, using symmetry arguments to combine into a single curve for each lattice spacing showing how the correlation in a given direction varies as the external in-plane field is rotated. Lines in right graphic are lines of best fit to Eqn. \ref{coseqn}.}
\label{DirectionalCorrHex}
\end{figure}

The hexagonal lattice is the sparsest of the lattices considered, which makes it the most difficult to accurately locate the islands in the array images. This leads to data with larger errors. While the data is not absolutely conclusive due to these errors, it is consistent with the observations on the square array that the main peak decreases with increasing lattice spacing, the overall minimum of the correlation decreases with increasing lattice spacing, and that there is a suppression of side peaks with decreasing interaction strength (see Fig. \ref{DirectionalCorrHex}) 

\newpage
\section{Kagome Lattice Data}

The kagome lattice is a decimated version of a sixfold rotationally symmetric lattice, in a frustrated geometry. Like the triangular lattice, it shows an increase in variation in correlation with increasing lattice spacing. Larger lattice spacings show an increase in next nearest neighbor correlations corresponding to the decreases in nearest neighbor correlations. 

\begin{figure} [ht]
\centering
\includegraphics[width = .7\textwidth]{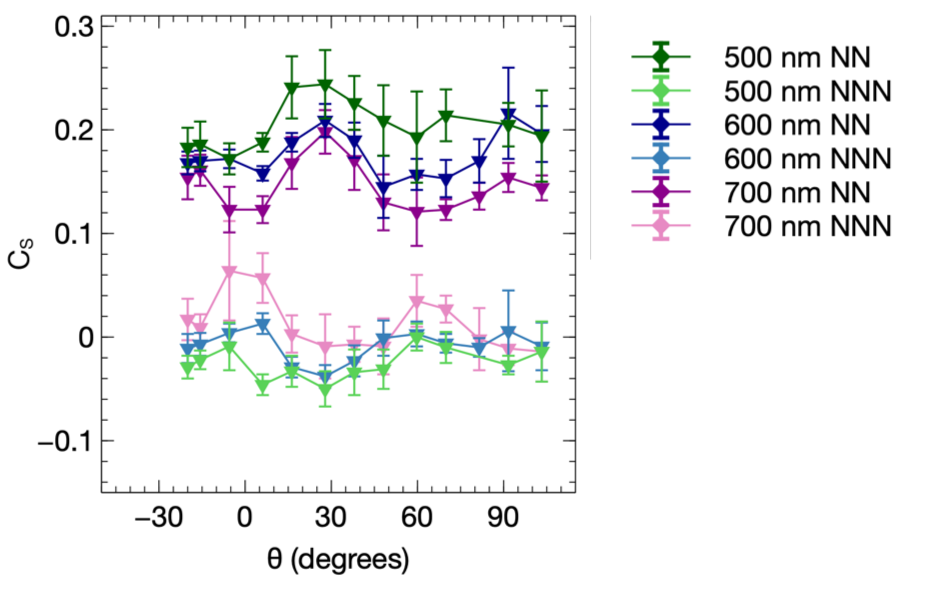}
\caption{$C_S$ values for hexagonal lattice with three different lattice spacings}
\label{Kagome}
\end{figure}

\begin{figure} [ht]
\centering
\includegraphics[width = .7\textwidth]{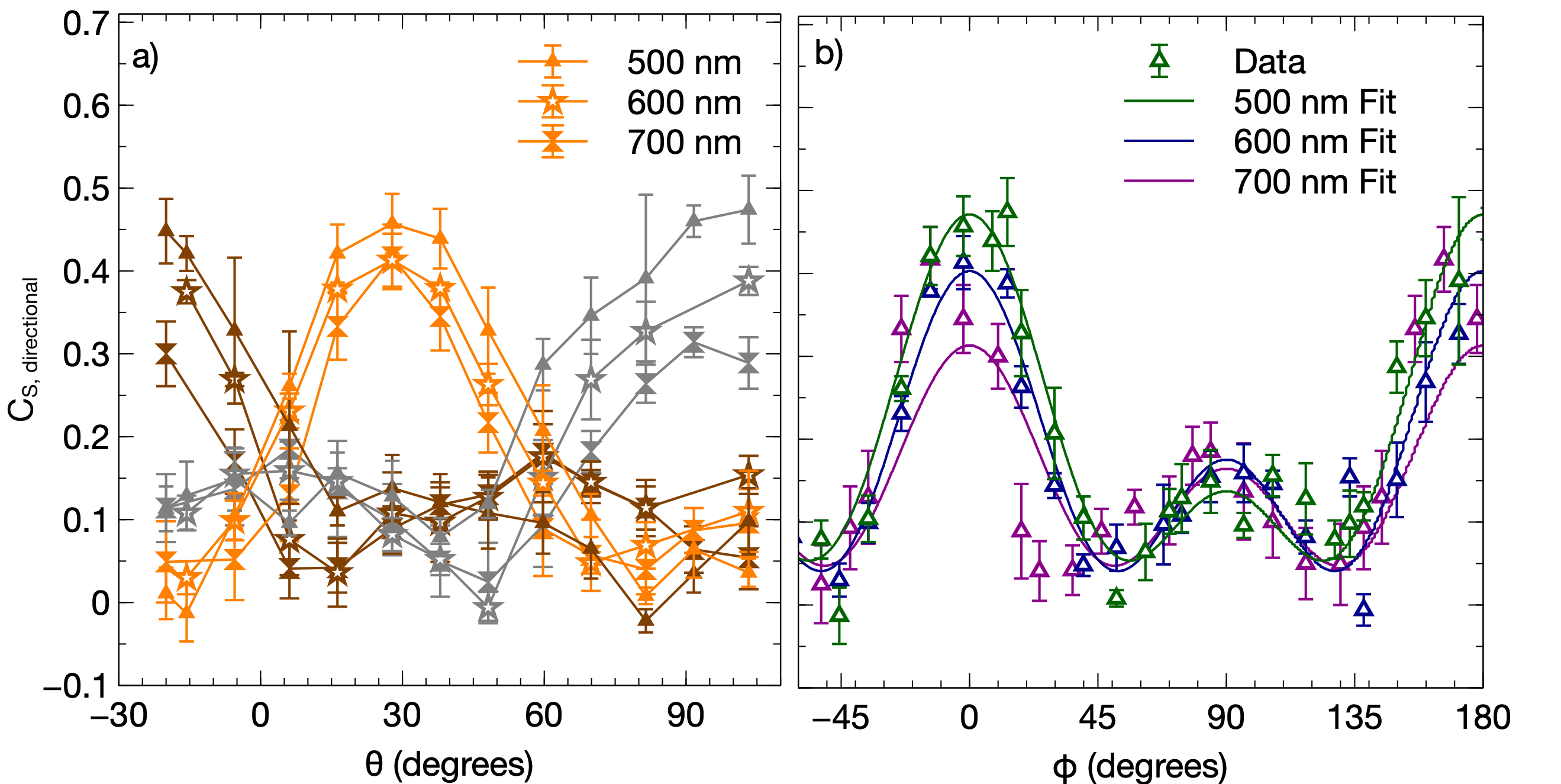}
\caption{a) Nearest neighbor correlation in kagome lattice calculated considering NN pairs oriented along D1, D2, and D3 as defined in the text as a function of $\theta$ for three different lattice spacings. b) The same data, using symmetry arguments to combine into a single curve for each lattice spacing showing how the correlation in a given direction varies as the external in-plane field is rotated. Lines in right graphic are lines of best fit to Eqn. \ref{coseqn}.}
\label{DirectionalCorrKag}
\end{figure}

The kagome lattice shows the same general features in the directional correlation measurements as the triangular data. The height of the main peak decreases steadily, the overall minimum correlation value is approximately zero, and the secondary peak is approximately constant in height as the interaction strength is decreased (see Fig. \ref{DirectionalCorrKag}). 

\newpage
\section{Fit Parameters}

Curve fitting to Eqn. \ref{coseqn} shown in Fig. \ref{DirectionalCorrTri}b and \ref{DirectionalCorrSqu}b in the text and \ref{DirectionalCorrHex} and \ref{DirectionalCorrKag} in the supplemental information is carried out using the curve\_fit function in the scipy.optimize package of Pyth\ref{DirectionalCorrSqu}b on, which optimizes the values of the fit parameters by minimizing the squared residuals. Standard deviation errors are calculated from the square root of the diagonal of the covariance array returned by the function.

\begin{table}
\begin{tabular}{|c|c|c|c|c|c|}
\hline
Lattice & Spacing (nm) & a & b & c & $\chi^2$ \\
\hline
\hline 
Hexagonal & 500 & .106 $\pm$ .009 & .129 $\pm$ .009 & .328 $\pm$ .006 & 0.73\\
\hline 
Hexagonal & 600 & .096 $\pm$ .009 & .133 $\pm$ .009 & .262 $\pm$ .006 & 0.94\\
\hline 
Hexagonal & 700 & .052 $\pm$ .011 & .102 $\pm$ .011 & .258 $\pm$ .008 & 1.31\\
\hline 
Kagome & 500 & .167 $\pm$ .008 & .109 $\pm$ .008 & .195 $\pm$ .006 & 1.22\\
\hline 
Kagome & 600 & .114 $\pm$ .006 & .117 $\pm$ .006 & .172 $\pm$ .003 & 1.01\\
\hline 
Kagome & 700 & .074 $\pm$ .011& .092 $\pm$ .011 & .147 $\pm$ .008 & 1.54\\
\hline 
Square & 500 & .131 $\pm$ .019 & .145 $\pm$ .017 & .326 $\pm$ .012 & 0.52\\
\hline 
Square & 600 & .093 $\pm$ .015 & .122 $\pm$ .014 & .218 $\pm$ .010 & 0.41\\
\hline 
Square & 700 & .067 $\pm$ .016 & .087 $\pm$ .015 & .139 $\pm$ .011 & 0.94\\
\hline 
Triangular & 500 & .218 $\pm$ .006 & .137 $\pm$ .006 & .208 $\pm$ .004 & 1.12\\
\hline 
Triangular & 600 & .123 $\pm$ .006 & .155 $\pm$ .006 & .169 $\pm$ .004 & 3.24\\
\hline 
Triangular & 700 & .080 $\pm$ .006 & .120 $\pm$ .006 & .125 $\pm$ .004 & 4.93\\
\hline 
\end{tabular}
\caption{Table showing the fitting parameters of the fits to directional correlation for hexagonal, kagome, square, and triangular lattices with spacings of 500 nm, 600 nm, and 700 nm, along with the corresponding $\chi^2$ values. \textit{a,b} and \textit{c} are as defined in Eqn. \ref{coseqn}. }
\end{table}

\end{document}